# Association Rule Based Flexible Machine Learning Module for Embedded System Platforms like Android

Amiraj Dhawan[1] Shruti Bhave[2]

Amrita Aurora[3] Vishwanathan Iyer[4]

*Abstract*—The past few years have seen a tremendous growth in the popularity of smartphones. As newer features continue to be added to smartphones to increase their utility, their significance will only increase in future. Combining machine learning with mobile computing can enable smartphones to become 'intelligent' devices, a feature which is hitherto unseen in them. Also, the combination of machine learning and context aware computing can enable smartphones to gauge users' requirements proactively, depending upon their environment and context. Accordingly, necessary services can be provided to users.

In this paper, we have explored the methods and applications of integrating machine learning and context aware computing on the Android platform, to provide higher utility to the users. To achieve this, we define a Machine Learning (ML) module which is incorporated in the basic Android architecture. Firstly, we have outlined two major functionalities that the ML module should provide. Then, we have presented three architectures, each of which incorporates the ML module at a different level in the Android architecture. The advantages and shortcomings of each of these architectures have been evaluated. Lastly, we have explained a few applications in which our proposed system can be incorporated such that their functionality is improved.

*Keywords*—machine learning; association rules; machine learning in embedded systems; android; ID3; Apriori; Max-Miner

## I. INTRODUCTION

Smartphones today are equipped with a number of features which have made it possible for users to obtain information at their fingertips. Incorporation of context aware computing in smartphones can give rise to innumerable new applications in mobile computing. A context aware system uses context to provide information and/or necessary services to users. The information and services provided depend on the current tasks the user is performing on the smartphone [1].

To utilize context awareness to its fullest potential, the system should have knowledge about how the device is used by the user and in what context. This can be achieved through machine learning. In machine learning, the system learns to make associations between the various tasks performed by the user and the corresponding context, which can be the inputs given to the device or other environmental factors. Android being a widely used mobile platform, in this paper, we have proposed methods to incorporate machine learning in the Android architecture. This is achieved through a machine learning (ML) module.

We have identified two major functionalities that the ML module should provide to achieve context awareness in the system and accordingly have described two modes of operation of the ML module. Next, we have proposed three variations in the Android architecture, which will enable machine learning to be included in the system. In each of these architectures, the ML module is placed at different levels in the Android architecture, which determines how the module will interact with the system as a whole. The placement of the ML module in the Android architecture determines which mode of operation it will be best suited for, as well as the applications that the ML module can be used for.

Finally, we have explained a few applications that can use context aware computing for a more user friendly smartphone experience. These applications make use of the ML module to learn about the device usage patterns and the context of the user. Accordingly, it forms certain associations and rules, using which these applications are prompted to the user proactively.

## II. LITERATURE SURVEY

### A. Android Architecture

Figure. 1 shows the principal components and levels in the Android architecture.

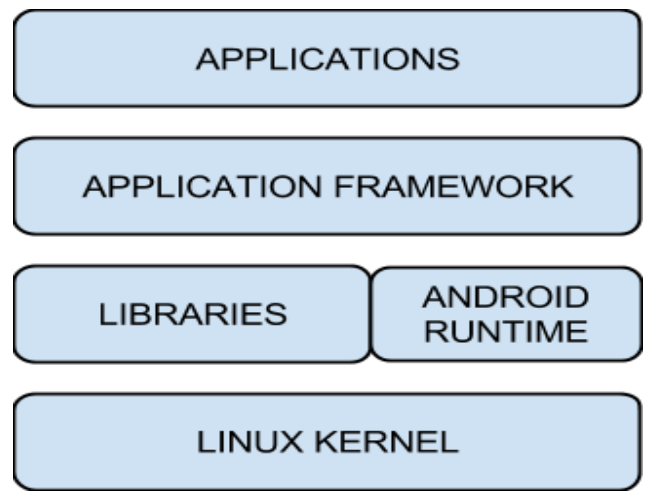

Fig. 1. The Android Architecture

The basic working and functionalities of each component are described as follows [2]:

- Linux Kernel:

Linux kernel forms the bottom layer of the architecture. It provides basic system functionality like process management,

device management and memory management. Also, the kernel handles functions like networking and a huge array of device drivers, to which peripheral hardware is interfaced.

- Libraries:

There is a set of libraries above the linux kernel, which includes WebKit- an open source web browser engine, libc, SQLite database-a useful repository for sharing and storage of application data, libraries to record and play audio as well as video, SSL libraries to monitor internet security, etc.

- Android Runtime:

This is the third section and belongs to the second layer from the bottom. It provides an important component named Dalvik Virtual Machine which is like Java Virtual Machine, designed and optimized specially for Android.

The Dalvik Virtual Machine uses core features of Linux like memory management and multi-threading. The Dalvik Virtual Machine enables every Android application to run in its own process, with its own instance of the Dalvik Virtual Machine. Additionally, the Android runtime provides a set of core libraries. Using these libraries, Android application developers can write Android applications using the standard Java programming language.

- Application Framework:

The Application Framework layer provides a number of higher level services to applications as Java classes. Application developers can use these services to write their applications.

- Applications:

All applications are strictly installed at the top layer. Home, Messages, Contact Books and Games are a few examples of these applications.

*B. ID3 (Iterative Dichotomiser 3) Algorithm*

ID3, developed by Ross Quinlan, is a decision tree learning algorithm used in the domains of machine learning and language processing. The ID3 algorithm employs a top-down search through the given datasets to test each attribute at every tree node. In this way, a decision tree is constructed. Every tuple in the database is then passed through the tree, which results in its classification.

To determine the input attribute that provides maximum information about the output, a metric called entropy is used. In information theory, entropy is used to measure the order or certainty in a given database set. A higher value of entropy indicates poor classification. A metric called information gain decides which input attribute is to be split. It is calculated as the difference between the entropy of the original dataset and the weighted sum of the entropies of the subdivided datasets. Information gain for each attribute is determined, and the attribute with the highest gain is split [3].

*C. Apriori Algorithm*

Apriori is a classical algorithm used for association rule learning. This algorithm tries to successively create larger and larger item sets that appear sufficiently in the database.

The algorithm starts with simple association rules of input to output mapping and tries to create more complex rules from these simple rules by increasing the size of the input and output item sets if they appear in the training data sufficient number of times. Every apriori algorithm requires two parameters, first is the support threshold and second is a confidence threshold. Apriori algorithm takes a bottom up approach since it starts with individual items in the item sets and successively adds multiple items in the item sets if they satisfy the support threshold.

This algorithm is one of the most popular association rule generation algorithms. But, the only disadvantage of this algorithm is in its bottom up approach which requires a lot of processing since it enumerates over all the combinations of the item sets [4].

*D. Max-Miner Algorithm*

Max-Miner algorithm is used to generate association rules with a complexity which is linear to the number of patterns or rules present in the training data.

The complexity is independent of the maximum length of these rules.The algorithm makes use of set-enumeration tree. It heuristically orders the items and dynamically reorders them on a pre-node basis, which leads to an improvement in performance. Hence, this algorithm is preferred [5].

*E. Papers*

In [6], how context aware systems enhance human computer interaction has been explained. Since context aware computing relies not only on the explicit current input given to a device, but also on the history of actions performed by the device and is capable of modifying the output with changing situations, the unnecessary interaction with the user is reduced.

In [7], the working of 'SenSay' has been described. SenSay is a mobile phone that is capable of manipulating its profile settings in accordance with the user's environment and physiological state. Various sensors, placed on the body give inputs about the users' state to a sensor box mounted on the waist. Based on the sensor information, a decision module computes the resultant action to be taken. However, SenSay does not incorporate machine learning, but relies solely on a set of predefined rules to determine the output. Also, SenSay gets inputs from sensors mounted on the body to obtain information about the user's context. Since our method does not require the user to carry any additional device on their person, it is more convenient to handle.

[8] explores the concept of using machine learning to train a system to choose songs to be played depending on the current activity and physiological state of the user. This device stores all the information relevant to a song, including the time of the day at which it is played, the corresponding activity of the user as well as the user's rating of the song. This is the training period of the system.

Once the training is complete, the device can select and play songs depending on the user's context using machine learning. However, its scope is limited only to song selection and music playlists.

### III. PROPOSED SYSTEM

#### A. Functionalities Required

Broadly, the proposed Machine Learning Module is supposed to provide two major functionalities as described below:

*1)   In the first case, any third party application should be able to use the module by providing the following information:*

*a) Set of inputs and/or parameters that are critical for the application.*

*b) Set of valid outputs/actions that the application understands.*

*c) Training data in the form of rules such as:*

**Inputs and/or Parameters =>Output/Action**, which are used to learn and automatically generate rules with their confidence level.

*2)   In the second case, the machine learning module works on its own in a global scope and tries to learn how and in what context the user uses the interface and the android system. Depending on the learning algorithm it should be able to automatically generate rules. Consider an example where the user frequently uses an application near a location (like a train station). Here, the system tries to learn this association. Accordingly, it generates a rule that if the user is near the same location, then the used application is invoked automatically.*

#### B. Modes of operation of ML module

In order to provide the above mentioned functionalities, the proposed machine learning module has two modes of operation as follows:

- Application Level Learning:

The module is used in this mode whenever other applications are to use the learning module for purposes internal to the application. This mode fulfills the first required functionality.

- System Level Learning:

The module constantly monitors the system and checks if the current context i.e., the current inputs and/or parameters match an already learned generated rule. If a match is found, then the required output/action is performed. Also, if an event occurs due to the users' intervention, like invocation of an application, the module adds the current inputs and/or parameters and the action i.e., the invocation of application as an entry in the training data. In this manner, the system learns new rules on the runtime.

#### C. Steps to be followed by the third party application in order to use the ML Module

To effectively use the ML module, any third party application needs to follow a set of steps to ensure proper configuration of the module and explain the kind of input and the expected list of outputs to the module. Since multiple applications should be able to use the ML module, a method is required through which the ML module can uniquely identify each application.

Accordingly, the module will load the proper context and use the inputs provided by the application. The process is as follows:

- Register_App:

In this step the third party application is required to call the Register_App API of the ML module in order to register itself with the ML module. This procedure generates a random alphanumeric string and stores this alphanumeric string along with the application name passed. This random alphanumeric string (identification key) is passed to the third party application which is required to remember this string. In any further communication with the ML module, this string is to be passed. The string helps the ML module to uniquely identify applications when the requests are coming through a common API.

In all the further requests, the string helps the ML module to load the proper context of the application for proper usage. This step needs to be done only once for initialization. After this step, the third party application is always required to send the identification key with the requests.

- Set_Input_Output:

In this step the third party application is required to send an array of parameters with their data types that are supposed to be used as inputs for the ML algorithm. It should also send an array of data types of the outputs which the ML module should produce, along with the identification key with which the applications are registered. This key helps the ML module to know which application is trying to access the API. This request is supposed to be sent only once, before calling Generate_Rules.

- Load_Training_Data:

In this step the third party can send any training data if available, to the ML module. This data should be in the format specified in the Set_Input_Output call. This step can be skipped and the data for generating the rules can be set as and when it is available as individual rows/tuples.

- Set_Training_Data_Row:

This step can be done repeatedly at any time or even skipped. This request is used to insert a new row in the training data for the applications, determined by the accompanying identification key. The format of the row should be as per the structure defined using Set_Input_Output.

Either Load_Training_Data or Set_Training_Data_Row should be performed (single or multiple times) to ensure that the ML module has some training data to generate rules. If the application does not have a training data set available then it can insert rows of training data whenever an event occurs. Hence this step is essential.

- Generate_Rules:

This step asks the ML module to generate the rules as per the support threshold and confidence threshold provided by the third party application. This request returns 'False' if the training data set is empty. If the training data for the application has at least one row, the returned data is the set of rules generated according to the support and confidence thresholds passed along with the request.  This request also stores the generated rules along with the identification key of the application for use. The design decision of allowing the application to provide the required support threshold and confidence threshold is to allow various applications to generate rules with a confidence level which is required by the application. This ensures that the generated rules are flexible according to the need of the application.

Consider a case in which an application does repeated calls to Generate_Rules with varying support and confidence threshold to smartly select which rule to follow and give some response to the event. This request is essential to be called at least once with success. This request can be configured in two modes:

*1) Automated:In this mode, the request is generated internally by the ML module whenever Set_Training_Data or Set_Training_Data_Row is called. The third party applications do not need to call this function directly*

*2) Manual: In this mode, the application is required to call this request explicitly whenever required.*

- Get_Current_Output:

The application after generating the rules from the training data can then use the rules internally to get the output for a given set of inputs. Else, it can ignore the returned rules and call this request Get_Current_Output and pass an array of the inputs to get the inferred output from the ML module. The module on receiving this request first loads the context of the application with the help of the identification key.

Next it loads the generated rules for this application and parses the rules to check if it has a rule with the inputs passed along with the request. If such a rule is found then the output of the rule is sent back to the application else'Null'is passed. The application, depending on the expected output sent by the ML module can then take actions for the event(s) (set of inputs passed to this request).

- Send_Feedback_Last_GCO_Request:

This stands for send feedback of last Get_Current_Ouput request. This request is used to send a feedback to the ML module for the last call to the request Get_Current_Output. If the output expected by the ML module was correct then a positive feedback is sent to the module else a negative feedback is sent. For a positive feedback, the module increases the confidence of the rule by a predefined amount. For a negative response the confidence of the rule is dropped by some predefined amount. This request is optional.

Some more non critical requests are as follows:

- Delete_Training_Data:

This request is used to delete all the training data.

- Delete_Training_Data_Row:

This request is used to delete a row from the training data. The set of inputs needs to be provided to find the row required for deleting. Either the first matched row is deleted or all the rows with the same values of the set of inputs are deleted.

- Change_Inputs_Outputs:

In case the application requires to change the structure of the training data then this request can be used to specify the new structure of the training data required. The missing inputs from the new structure are deleted from the training data and any new input is kept as null for the previous training data. Any changes in the output are also handled the same way as inputs.

### D. System Level Learning: A special case of Application Level Learning

System level learning is a special case of application level learning in which the entire system is one application and the requests are auto generated by listeners on the sensors (inputs/parameters) and events (outputs like vibrator or speakers, invocation of an application). If there is an event like invocation of an application by the user, the state of the sensors with some other parameters like time etc, are used as the input. The current inputs are monitored constantly. At any point if the inputs match a rule, the output as per the rule can be used to automatically generate an event without the users' intervention.

### E. Architectures

Placing the Machine Learning Module in the android architecture is a critical decision to be made. The placement affects the way other applications would interact with the ML module and also how the module can work independently in the System Level Learning mode.

- Application Level Architecture

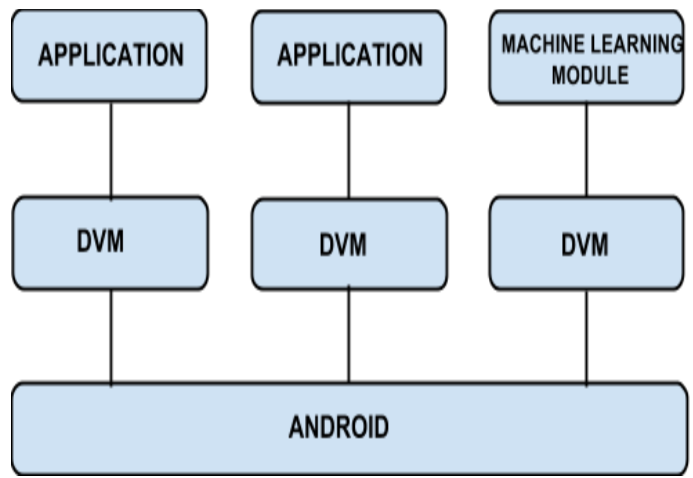

Fig. 2.   Application level architecture

In this proposed architecture, the machine learning module is used as an application which executes over its own Dalvik Virtual Machine (DVM). The third party applications would be required to communicate with the Machine Learning Module Application (MLMA). This can be achieved using the Intent class of the Android Framework for Inter Application Communication.

The following snippet shows how inter application communication can be achieved.

```
<activity android:name=".SecondActivity">
<intent-filter>
<action
android:name="com.machineLearning.action.REGISTER_AP
P_ML" />
<action
android:name="com.machineLearning.action.SET_INPUTS_
OUTPUTS" />
<action
android:name="com.machineLearning.action.GET_CURREN
T_OUTPUT" />
/* and so on for all the requests mentioned above */
</intent-filter>
</activity>
```

This method requires the third party applications to specify an action for each of the functions provided by the MLMA. This approach is not preferable since it requires a lot of effort from the third party application. Also in this approach since the ML module runs as an independent application, it cannot access all the various sensors and input data sources without proper permissions. Thus in this case the ML module would require access to all the sensors and in return permissions to all the sensors which is difficult to manage.

In this case since the ML module works as a separate application, using the module for system wide learning is difficult. For system level learning the operating system would be required to use the functionality of the module as if it is another third party application. Thus system level learning cannot work independently without the need for the operating system to use it.

- System Level Architecture

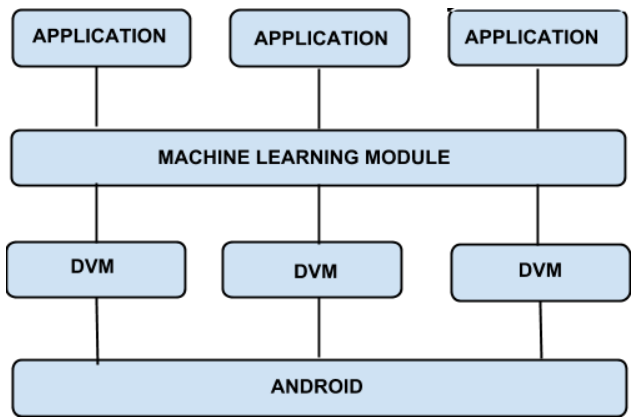

Fig. 3. System Level Architecture

In this approach, the ML module works as a layer between the DVM and the applications. This gives easy access to the machine learning module from the applications.

In this case since the ML module does not act as a standard application, it can be bundled with the operating system itself and surpass the permissions required as in the previous approach.

The module being part of the operating system can easily access any sensor. Since in this approach the ML module is between the DVM and the third party applications, it can easily monitor the usage of the applications and can work well for system level learning. However since the ML module is shared across the third party applications, providing access to the module from the applications for application level learning is challenging. Separating the applications context from each other would be difficult since all the applications share the same ML module layer.

This approach would require a lot of change in how the applications interact with the DVM instances. The module would be required to monitor all the interactions between the applications and the corresponding DVM machine. This would require changes in the core android operating system.

- Hybrid Architecture

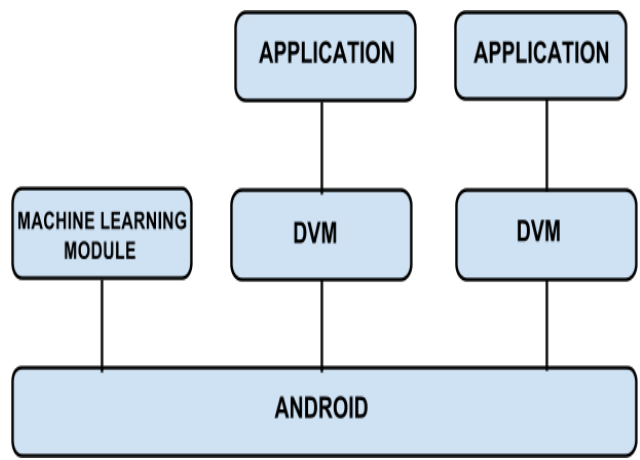

Fig. 4. Hybrid Architecture

In this approach, the module works as an independent system level library which is available as an API for applications for usage. Since the module is an independent library, the operating system too can use the functionality and perform system level learning. This approach can be used effectively for both system level as well as application level learning. Since the ML module is a library, it does not require permissions to access the sensors or other input sources. The applications can communicate with the module using all the requests explained earlier.

The operating system can register itself and send training data to learn usage patterns. Thus, it can deduce which application is highly likely to be used at the current instance. This output or deduced information can be used to automatically start applications as per the current input and sensor state. This architecture not only provides a good and clean interface for application level learning but also supports system level learning.

### F. Generation of Rules

Once the application (or operating system in case of system level learning) creates a training data set for the ML module using either Load_Training_Data or Set_Training_Data_Row, it can instruct the machine learning module to generate association rules. These can be used to predict the current usage output as per the current input state. This procedure may take considerable time depending on the size of the training data. The module at this point starts parsing the training data for the particular application and tries to generate association rules that cross the support and confidence threshold values. The ML module generates these association rules by using algorithms like Apriori, ID3 or Max-Miner, which are popular association rule generation algorithms. For this application Max-Miner algorithm would be preferable since it uses less resources to come up with good association rules as compared to Apriori algorithm. Since embedded systems have lesser resources, performance should be the most important criteria while selecting the association rule generation algorithm.

Once the rules are generated for an application, it can then request for the current output possible according to the rules generated by providing the input list. If a rule exists with similar input state as provided by the application, the corresponding output is returned by the machine learning module along with its confidence level. The application can then decide what is supposed to be done with the output.

In case of the system level learning, the operating system performs all the tasks just like an ordinary application can work.

The actual end action is decided by the applications depending on the output of the machine learning module. This output can also be ignored by the application thus proving flexible usage of the module.

## IV. Applications

The following examples provide an overview of the varied use of the machine learning module in mobile applications. Depending on the users everyday usage of the device, the machine learning module trains itself and provides an output that saves time and is comfortable to access by the user.

### A. Music player application

A user has an everyday schedule of listening to particular songs at particular instants of time, for example, while travelling to work at 8:00 AM or returning home at 7:00 PM. The machine learning module stores all the information relevant to opening of the music player and playing the song.

This includes the time at which it is played and the activity of the user at that time. Inputs of inserting the headphones in the device and opening the player at that particular time are provided to the module. It then trains itself according to this users schedule and generates an application ID which denotes that the music player application is running currently. If the same task is performed by the user every day at the same time after the training period, the module automatically provides a widget to the user at that time of the day to open the player thus saving his time.

### B. Automatic Profile Settings

Automatic profile settings can be extremely beneficial to students in schools and colleges as well as employees in companies. The predefined time of a daily lecture or meeting is known to the user. Inputs of these lecture/meeting timings as well as the location of the school/office is provided as input to the machine learning module. Combining the users' environmental location and the timing of the users usage on the mobile device, the module trains itself and depending on the application ID provided, it displays a widget suggesting to the user to change the phone settings to silent mode. This prevents the unnecessary ringing of mobiles in important meetings and lectures making it immensely user friendly.

### C. Task List

The user experience can be drastically improved by displaying relevant information on the device's screen rather than spending time searching for that information. Machine learning module takes input of a task list and the time for execution of this task from the user. Following the users usage pattern, it trains itself to display a widget of this task list needed at the time of its implementation. For example, if the user purchases groceries from the market every Friday at 7:00 PM, then the mobile will automatically display the list of groceries to be purchased by the user at this time every week. This saves time and makes it user friendly for the user.

### D. Messaging

The user habitually sends the same alert message at a particular instant every day at the same time for a week. For example, a user before leaving for school/work informs his parents/spouse that he has left home. The ML module saves a draft of this message. The time, day and message are provided as the input to the ML module. Depending on the time span when the user leaves for school/work, the module displays an output i.e., the drafted message on the users screen with the recipients entered. The user can then just press the send option to send the message thus saving his time rather than retyping the same message every day and then searching for the recipients.

### E. Location Based Profile Settings

If the user traverses the same path every day eg: from home to work or vice versa, the ML module keeps a record of his path taken regularly which is provided as input. With the help of Google plus, if there is a huge amount of traffic on that path on any particular day, it informs the user to traverse another path with lesser traffic. Thus the users' energy and time are saved.

### F. Alarm

The user sets the alarm every day in a specific span of time. Thus the time and date of entering the event is provided to the ML module as input. The ML module then learns and trains itself according to the input of these events. The module provides an application ID which indicates the application being that of the alarm. The output is provided to the user in form of a widget that suggests the needed occurrence of the alarm. Thus the module trains itself to display a widget of the alarm every day at certain time interval before the person sets the alarm.

## V. Conclusion

Incorporation of context awareness in mobile computing has a wide scope in a number of smartphone applications. The ability to learn about users' preferences and usage patterns and suggest services accordingly will facilitate a more user-friendly smartphone experience.

Although this paper describes context awareness and machine learning specific to the android framework, it can also be extended to include other embedded systems like set-top box or a set-top unit. The machine learning module can learn about the channel preferences of the viewer, on the basis of the previous viewing history, time as well as day of the week. Accordingly, it can suggest which channel is to be played at what time. As most viewers have fixed television schedules as well as fixed preferences in terms of channels and television shows, the channels can be tailored according to each user's needs and demands. Thus, the machine learning module can be incorporated in various embedded systems for a more personalized and simplified usage experience.

## VI. FUTURE WORK

The paper in its current scope focuses on the importance of a machine learning module in an embedded system platform like android. The main focus is on how to design the module to be flexible so that it can be used by other applications as well as by the system. It also deals with the interface of the module to allow maximum flexibility and allow efficient use of the module.

Taking this proposed system as the base, future efforts include generalizing the module for generic embedded system platforms, analyze and compare the performance of the module using various association rule generation algorithms like ID3, Apriori and Max-Miner from a qualitative perspective. The future work on this system also includes working on the privacy issues of this machine learning module since it handles user data. Another research area could be to design and implement the machine learning module in a hardware integrated circuit to offload the heavy processing required by the module from the main processing unit. The hardware IC can be designed to support processors based on the ARM architecture which is widely used on embedded systems. This could lead to better support for machine learning on such systems and provide unparalleled support to understand the user/environment better in order to take smart decisions by the embedded platforms.

Robotic platforms can leverage the machine learning capabilities to be slightly closer to achieving artificial intelligence. This is because these platforms can then accurately predict future events based on past events. If the module is implemented as a hardware IC, then the interface between the robotic platforms and this module can be simplified.